\documentclass[preprint,showpacs,amsmath,amssymb]{revtex4}
\usepackage{graphicx,amsmath,amssymb,bbm,times}
\begin{document}
\title{Light statistics by non-calibrated linear photodetectors}
\author{Maria Bondani} \email{maria.bondani@uninsubria.it}
\affiliation{National Laboratory for Ultrafast and Ultraintense
Optical Science - C.N.R.-I.N.F.M., Como, Italy}
\author{Alessia Allevi}
\affiliation{Consorzio Nazionale Interuniversitario per le Scienze fisiche della Materia, C.N.I.S.M., Universit\`a degli Studi dell'Insubria, Como, Italy}
\author{Alessandra Andreoni}\affiliation{C.N.I.S.M. and Dipartimento di Fisica e
Matematica, Universit\`a degli Studi dell'Insubria, Como, Italy}
\date{\today}
\begin{abstract}

We theoretically demonstrate that detectors endowed with internal gain and operated in regimes in which they do not necessarily behave as photon-counters, but still ensure linear input/output responses, can allow a self-consistent characterization of the statistics of the number of detected photons without need of knowing their gain. We present experiments performed with a photo-emissive hybrid detector on a number of classical fields endowed with non-trivial statistics and show that the method works for both microscopic and mesoscopic photon numbers. The obtained detected-photon probability distributions agree with those expected for the photon numbers, which are also reconstructed by an independent method.
\end{abstract}
\maketitle
\section*{INTRODUCTION}

Full characterization of quasi-monochromatic electromagnetic fields requires the knowledge of amplitudes and phases, which can be achieved by using homodyning detection techniques. These techniques, which belong to routine in radio-frequency analysis, were brought to operation at optical frequencies \cite{yuen1980} and became popular since the discovery of phenomena generating fields with non-classical properties (see Ref. \cite{zavatta2006} for a review). They were used not only in the case of quasi-monochromatic light \cite{smithey1993}, but also in the case of definitely broadband fields, that is of pulsed fields \cite{grangier2004}. However, a simple information concerning the field amplitude such as a statistical information, for instance of the photon-number, is often sufficient and sometime crucial to understand light-matter interaction phenomena \cite{mukamel2003}.

It might be thought that measuring photon-number statistics is feasible only when very few photons are present in the field to be characterized \cite{mandel1995}. This prejudice stems from the notably poor photon-number discriminating capability of detectors, which hardly goes beyond five detected photons \cite{burle, Hamamatsu} for detectors based on the photoelectric effect. Many efforts have been made to achieve better photon-counting capabilities by working both on the detectors and on the front-end optics. Multi-pixel detectors with single-photon sensitivity in each pixel, such as arrays of single-photon solid-state detectors \cite{KOKbraunstein}, intensified CCD cameras \cite{iCCD} and silicon photomultipliers \cite{SiPM}, were (are being) shown to allow photon counting provided the light is spread across the sensitive area so that each pixel receives one photon at most. Photons temporally spread by either cascaded arrays of beam splitters or multiple fiber-loop splitters have been alternatively used in connection with single-photon avalanche diodes \cite{Tmultiplex}.
Some of us showed that an on/off detector, that is a detector giving a standard current output as the response to any $n\ge 1$ photon number, allows determining the whole statistical distribution of photons from the experimental probability of detecting zero photons at varying the detection efficiency. The technique, based on the application of a maximum likelihood algorithm, has been exploited to reconstruct both classical and quantum states \cite{zambra2005, zambra2007}. However, the method loses efficacy at increasing $n$, nor the result improves if the detector has photon-number discriminating capabilities for $n\geq 1$ \cite{zambraSOLO}.

In this work we use a hybrid photo-detector (HPD), which is a photoemissive detector endowed with internal gain as it includes a photocathode and an avalanche diode. The pulse-height spectra of the output charge of this HPD exhibit separated peaks typically for $m\leq 3-4$ photoelectrons. We perform single-shot measurements on ps-pulsed classical fields producing mean charge values from below to well above such $m$ values. From the converted voltages, $v$, collected for thousands of measurements, we determine the ingredients of the Fano factor of the voltages, {\it i.e.} mean and variance of the $P_v$ distribution. We show that, if the detector is operated within its range of linear input/output response, we can retrieve $P_{m}^{\ el}$ from the experimental $v$-data without need of calibrating either the HPD or the electronics that processes its output. In the cases of relatively low intensities, in which the $P_{0}^{\ el}$ peak is sizeable, we apply the algorithm cited above \cite{zambra2005} to reconstruct the statistical distribution of photons and find $P_{n}^{\ ph}$ distributions in good agreement with the theoretical ones. It is noticeable that the method easily allows determining the statistical distribution of $n\le 100$ photons, that is in the mesoscopic regime, which has been scarcely explored for the above reasons.

The paper is organized as follow. In Section~\ref{sec:teo}, we develop the theory that allows the retrieval of the detected-photon distribution, $P_{m}^{\ el}$, from the experimental $P_{v}$. In Section~\ref{sec:exp}, we check the theoretical results with experiments performed on a number of non-trivial classical fields with our hybrid photo-detector. For each type of field, in Section~\ref{sec:onoff} we reconstruct the distributions of the photon-numbers $P_{n}^{\ ph}$ from the simple knowledge of the $P_{0}^{\ el}$ peaks of the distributions retrieved in Section~\ref{sec:exp}. In the conclusive Section we present some perspectives of our method as one that allows determining photon-number statistics with linear non-photon-counting detectors.

\section{THEORY}\label{sec:teo}

For any photo-detection experiment performed with an overall detection efficiency $\eta<1$, the distribution of the number of electrons generated by the primary photon interaction that occurs in the detector, $P_{m}^{\ el}$, is linked to that of the number of photons in the light under measurement, $P_{n}^{\ ph}$, by \cite{mandel1995, agliati2005, zambra2005}
\begin{equation}
 P_{m}^{\ el}=\sum_{n=m}^{\infty} \left(
 \begin{array}{c}n\\m\end{array}\right)
 \eta^m (1-\eta)^{n-m} P_{n}^{\ ph}\ .\label{eq:phel}
\end{equation}
If we limit our analysis to the first two moments of the distributions, the link between the statistics of photons and that of the detected photons is given by \cite{agliati2005}
\begin{eqnarray}
 \bar{m} &=& \eta \bar{n}\ \ ; \ \ \sigma_{m}^{(2)} = \eta^2\sigma_{n}^{(2)}+\eta(1-\eta)\bar{n}\label{eq:momentsPHEL}\ ,
\end{eqnarray}
in which $\bar{n}=\sum_{n=0}^{\infty} nP_{n}^{\ ph}$ is the mean value and $\sigma_{n}^{(2)}=\sum_{n=0}^{\infty} (n-\bar{n})^2 P_{n}^{\ ph} $ is the variance. The same notation is adopted for $\bar{m}$ and $\sigma_{m}^{(2)}$.

In principle, $ P_{m}^{\ el}\neq  P_{n}^{\ ph}$, that is the distribution of the photoelectrons is not necessarily the same as that of the photons. As a matter of fact, all the fields studied in this work are endowed with a photon-number distribution that is invariant under the convolution in Eq.~(\ref{eq:phel}). We will exploit this property to compare the experimental results with the corresponding theoretical distributions.

For photoemissive detectors, in which the primary detection process is the photoelectric effect, $m$ is the number of photoelectrons emitted by the photocathode, which normally undergo multiplication before reaching the anode. For the linearity of the response, both the amplification internal to the detector (gain) and that of the electronics that processes its output, must be independent of $m$. In the following we also adopt the strongly simplifying hypothesis that the spread of the single photoelectron peak in the final electronic output is negligible as compared to its mean value. Under such an hypothesis, which is too weak to ensure photoelectron counting capability, the relation linking the statistics of the photoelectrons to that of the final voltage outputs, $v$, is
\begin{equation}
 P_{v} = \frac{1}{\gamma} P_{\gamma m}^{\ el}\ ,\label{eq:outs}
\end{equation}
being $\gamma$ a conversion coefficient describing the overall $m$-to-$v$ conversion process. The distribution in Eq.~(\ref{eq:outs}) is the experimental $P_v$. Its first two moments are given by
\cite{agliati2005}
\begin{eqnarray}
 \bar{v}=\gamma \bar{m}\ \ ;\ \ \sigma_{v}^{(2)} = \gamma^2 \sigma_{m}^{(2)}\ \ ,\label{eq:momentsVOUT}
\end{eqnarray}
where the symbols are defined as in Eq.~(\ref{eq:momentsPHEL}).
\par
Apparently Eq.~(\ref{eq:outs}) allows retrieving $P_{m}^{\ el}$ from the experimental $P_{v}$ only if $\gamma$ is known. The problem is that to obtain the $\gamma$-value one should repeat the measurement of $P_{v}$ for the same state of light at different mean numbers of photons $\bar{n}$ by using a detector with photoelectron counting capability. We show that we can avoid such direct calibrations and determine the $\gamma$-value by other means. In fact, if we calculate the Fano factor for the output voltages, $F_v$ , by using Eq.~(\ref{eq:momentsPHEL}) and Eq.~(\ref{eq:momentsVOUT}) we find
\begin{eqnarray}
 F_v &\equiv& \frac{\sigma_{v}^{(2)}}{\bar{v}} = \frac{{\gamma ^2 [
 {\eta ^2 \sigma_{n}^{(2)}  + \eta (1 - \eta)\bar{n}}]}}{{\gamma \eta \bar{n}}}\nonumber\\
 &=& \gamma \eta F + \gamma(1 - \eta)=\gamma \eta Q + \gamma\ ,
\label{eq:fano}
\end{eqnarray}
where we have rewritten the Fano-factor for the photons, $F=\sigma_{ph}^{(2)}(n)/\bar{n}$, as $F=1+Q$ by inserting the Mandel $Q$-factor \cite{mandel1979}. By multiplying and dividing Eq.~(\ref{eq:fano}) by $\bar{n}$ and reusing the above equations we get
\begin{equation}
 F_v  = \frac{Q}{\bar{n}} {\bar{v}} + \gamma\ ,
\label{eq:fanoQ}
\end{equation}
in which $Q/\bar{n}=(\sigma_{ph}^{(2)}(n)-\bar{n})/\bar{n}^2$ depends only on the specific state under measurement. Thus Eq.~(\ref{eq:fanoQ}) shows a linear dependence of the $F_v$-factor on $\bar{v}$, with $Q/\bar{n}$ as the proportionality coefficient, which is zero for Poissonian light, positive for classical super-Poissonian light and negative for nonclassical sub-Poissonian light. This linearity can be verified by repeatedly measuring a field upon inserting filters with different transmittance in front of the detector. In fact the insertion of filters amounts to varying $\bar{v}$, but leaves unaltered the quantity $Q/\bar{n}$. In practice, as the expression in Eq.~(\ref{eq:fanoQ}) is general, we expect that, for any given statistics, upon plotting the experimental values of $F_v$ as a function of $\bar{v}$ the data points should align along a straight line with intercept $\gamma$ and slope $Q/\bar{n}$.

\begin{figure}[h]
\includegraphics[width=40mm,angle=270]{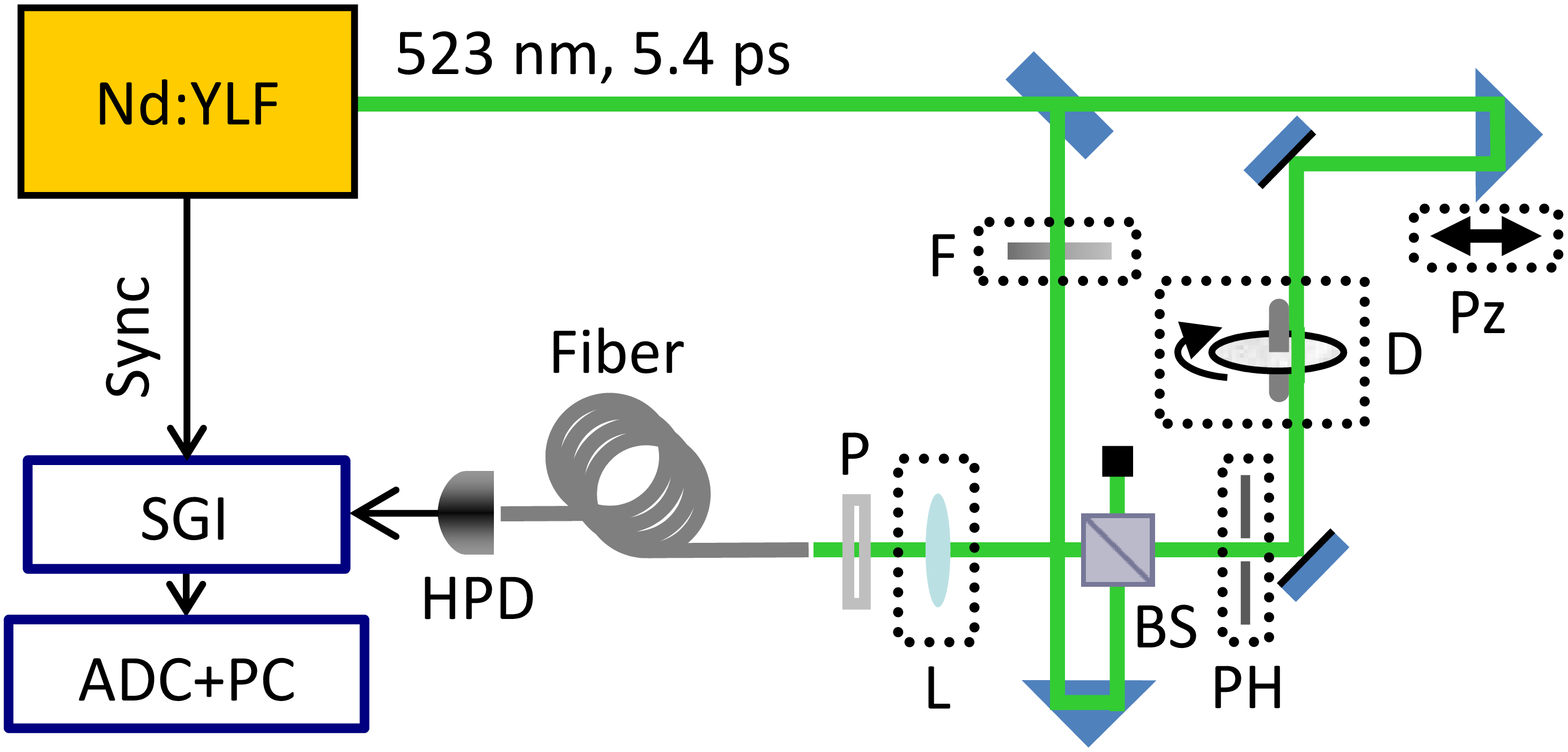}
\caption{(Color online) Schematic diagram of the experimental setup. HPD, hybrid photo-detector; SGI, synchronous gated integrator; F, neutral density filter; BS, 50$\%$ beam splitter; L, focusing lens; P, polarizer; Pz, piezoelectric movement; D, diffuser (rotating ground glass plate); PH, pin-hole. Components in dotted boxes are inserted/activated when necessary.}\label{f:scheme}
\end{figure}

In order to exploit Eq.~(\ref{eq:fanoQ}) to retrieve the $P_m^{\ el}$ distributions for lights with different photon-number statistics, we built some "artificial" field states by mixing, at a beam splitter, different fields generated as described in Section~\ref{sec:exp}. With reference to Fig.~\ref{f:scheme} we anticipate a summary of the schemes we adopted to produce the field states to be measured: (A) coherent light directly from the laser source; (B) single-mode thermal light obtained by passing the coherent field through a diffuser, namely a rotating ground glass plate (D in Fig.~\ref{f:scheme}) and selecting a single speckle by a pin-hole (PH in Fig.~\ref{f:scheme}); (C) multi-mode thermal light obtained as in (B), where multiple speckles where selected by a wider pin-hole and focused by lens L; (D) phase-averaged displaced coherent light obtained by mixing two coherent fields at beam splitter BS upon averaging the relative phase between them by using a piezoelectric movement (Pz in Fig.~\ref{f:scheme}); (E) displaced single-mode thermal light obtained by mixing the state generated in (B) with a coherent field; (F) displaced multi-mode thermal light (same as (E) with state (C) instead of (B)). In cases (A), (B) and (C), the neutral-density filter F in  Fig.~\ref{f:scheme} is substituted by a beam stop. In cases (D), (E) and (F), F is inserted on the coherent beam. As states (D), (E) and (F) are characterized by non-trivial statistics, reconstructing $P_{m}^{\ el}$ for such states would be a rather convincing test of the goodness of our method.

In the following we present the expressions of the statistical distributions, $P_{n}$, of the photons for the above states, for which we calculate the slope coefficients $Q/\bar{n}$ of $F_v$ as a function of $\bar{v}$ in Eq.~(\ref{eq:fanoQ}). All mean photon numbers appearing in the following calculations are taken at the BS output.

\subsection{Coherent}

The photon-number distribution of the coherent field is given by the Poisson distribution
\begin{equation}
 P_{n}  = \frac{|\alpha|^2}{n!} e^{-|\alpha|^2}\ ,\label{eq:poiss}
\end{equation}
for which $\bar{n}=\sigma_{n}^{(2)}=|\alpha|^2$, so that, according to Eq.~(\ref{eq:fanoQ}):
\begin{eqnarray}
 F_v &=& \gamma  \label{eq:fanoCOH}
\end{eqnarray}
independently of the mean value $\bar{v}$.

\subsection{Single-mode thermal}

The photon-number distribution of a single-mode thermal field is given by
\begin{equation}
 P_{n} = \frac{n_{th}^n}
 {\left(n_{th}+1\right)^{n+1}}\ , \label{eq:term}
\end{equation}
for which $\bar{n}=n_{th}$ and $\sigma^{2}_{n}= n_{th}\left(n_{th}+1 \right)$. In this case, Eq.~(\ref{eq:fanoQ}) gives
\begin{equation}
 F_v = \bar{v}+\gamma\ . \label{eq:fanoTHER}
\end{equation}

\subsection{Multi-mode thermal}

For $\mu$ independent thermal modes equally populated by $N_{th}/\mu$ mean photons \cite{mandel1995} we have the distribution
\begin{equation}
 P_{n}  = \frac{\left(n +\mu-1\right)!}
 {n!\left(\mu - 1 \right)! \left(N_{th}/\mu+1 \right)^{\mu} \left(\mu/N_{th}+1 \right)^{n}}\ , \label{eq:multit}
\end{equation}
for which $\bar{n}=N_{th}$ and $\sigma^{2}_{n}= N_{th}\left(N_{th}/\mu+1 \right)$. Equation~(\ref{eq:fanoQ}) gives
\begin{equation}
 F_v = \frac{\bar{v}}{\mu}+\gamma
 \ . \label{eq:fanoMTHER}
\end{equation}
Note that the slope coefficient $1/\mu$ has a value lying between 0 and 1.

\subsection{Phase-averaged displaced coherent}

The state is the superimposition of a coherent state with $|\alpha_1|^2$ mean photons and a phase-averaged state with $|\alpha_2|^2$ mean photons at the same frequency. The expected photon-number distribution for such a state is given by \cite{zambra2007}
\begin{equation}
 P_{n} = \frac{A^{n}}{n!}e^{-A}\sum_{k=0}^n \left(
 \begin{array}{c}
   n  \\
   k  \\
 \end{array} \right)
 \frac{\left(-1\right)^k }{2\pi}\left(\frac{B}{A}\right)^k\frac{\Gamma\left(1/2+ k/2\right)
 \Gamma\left(1/2\right)}{\Gamma\left(1+ k/2\right)}
 {}_1F_2 \left[\left\{1/2+k/2\right\},\left\{1/2,1+k/2\right\},B^2/4 \right]\ , \label{eq:phaseaver}
\end{equation}
where $A = |\alpha_1|^2+|\alpha_2|^2$, $B=2 |\alpha_1||\alpha_2|$ and $_1F_2(a,b,z)$ is the generalized hypergeometric function.
The distribution in Eq.~(\ref{eq:phaseaver}) has mean $\bar{n}=|\alpha_1|^2+|\alpha_2|^2$ and variance $\sigma^{(2)}_{n}= \bar{n}\left(K\bar{n}+1 \right)$, with $K\equiv 2|\alpha_1|^2|\alpha_2|^2/(|\alpha_1|^2+|\alpha_2|^2)^2$. We thus have:
\begin{equation}
 F_v = K \bar{v}+\gamma \ , \label{eq:fanoAVER}
\end{equation}
in which the value of the slope, $K$, lies between 0 and 1/2. This fact makes the dependence on $\bar{v}$ in Eq.(\ref{eq:fanoAVER}) not distinguishable from that in Eq.~(\ref{eq:fanoMTHER}) when $\mu>2$.

\subsection{Displaced thermal}

The state is the superimposition of a thermal state having $n_{th}$ mean photons and a coherent state with $|\alpha|^2$ mean photons. The photon-number distribution is
\begin{equation}
 P_{n}  = \frac{n_{th}^n}
 {\left(n_{th}+1\right)^{n+1}}\exp\left(-\frac{|\alpha|^2}
 {n_{th}+1}\right) L_n\left(-\frac{|\alpha|^2}
 {n_{th}\left(n_{th}+1\right)}\right) , \label{eq:DISPLterm}
\end{equation}
in which $L_n$ is the Laguerre polynomial $L_n^\gamma$ for $\gamma=0$. In this case we find $\bar{n}=n_{th}+|\alpha|^2$, $\sigma^{(2)}_{n}= n_{th}+|\alpha|^2+ n_{th}\left(n_{th}+2|\alpha|^2\right)$ and finally:
\begin{equation}
 F_v = K_1 \bar{v}+\gamma \ , \label{eq:fanoDISPLter}
\end{equation}
where $K_1 = n_{th}\left(n_{th}+ 2|\alpha|^2\right)/\left(n_{th}+|\alpha|^2\right)^2$.
Note that the value of $K_1$ lies between 0 and 1 on varying $|\alpha|^2$.

\subsection{Displaced multi-mode thermal}

The state is the superimposition of the state in subsection~(C) with a coherent state with $|\alpha|^2$ mean photons. The photon-number distribution is given by
\begin{equation}
 P_{n}  = \frac{N_{th}/\mu^n}
 {\left(N_{th}/\mu+1\right)^{n+\mu}}\exp\left(-\mu\frac{|\alpha|^2}
 {N_{th}/\mu+1}\right) L_n^{\mu-1}\left(-\mu\frac{|\alpha|^2}
 {N_{th}/\mu\left(N_{th}/\mu+1\right)}\right) , \label{eq:DISPLmterm}
\end{equation}
for which we have $\bar{n}=N_{th}+\mu|\alpha|^2$ and $\sigma^{(2)}_{n}= N_{th}+\mu|\alpha|^2+ N_{th}\left(N_{th}/\mu+2|\alpha|^2\right)$. Thus we find:
\begin{equation}
 F_v = K_2 \bar{v}+\gamma\ , \label{eq:fanoDISPLmter}
\end{equation}
where $K_2 = K'_1/\mu$, being $K'_1$ defined as $K'_1 = N_{th}/\mu\left(N_{th}/\mu+ 2|\alpha|^2\right)/\left(N_{th}/\mu+|\alpha|^2\right)^2$ in analogy with case (E). In this case, on varying $|\alpha|^2$, the slope $K_2$ takes values between 0 and $1/\mu$.

Note that the slopes $K$ and $K_1$ in the above cases (D) and (E), respectively, can be simply expressed in terms of the ratio of the intensities of the mixed fields, while for $K_2$ the knowledge of the number of modes $\mu$ is also necessary.
We finally remind that all the above distributions are identical to those of photoelectrons, upon rescaling the parameters by the overall detection efficiency, $\eta$.

\section{EXPERIMENTAL RESULTS AND DISCUSSION}\label{sec:exp}

All the measurements presented here were performed on picosecond-pulsed light fields at 523 nm wavelength. The light source was the second-harmonic output of a Nd:YLF mode-locked laser amplified at 5 kHz (High Q Laser Production, Austria) producing pulses of $\sim$5.4~ps duration.
We used a multimode fiber with 600~$\mu$m core diameter to deliver the light to the HPD detector (H8236-40, Hamamatsu, Japan, maximum quantum efficiency of the photocathode: 0.40 at $\sim$550 nm). The maximum overall detection efficiency, calculated by including the losses of the collection optics, was $\eta_{max}=0.29$. The detector was strictly operated within its range of linear response. Its output current pulses were suitably gate-integrated by SR250 modules (Stanford Research Systems, CA) and sampled to produce a voltage, $v$, which was digitized and recorded at each shot.
\par\noindent
As sketched in Fig.~\ref{f:scheme}, we used a 50$\%$ non-polarizing beam splitter, BS, to mix the fields in cases (D), (E) and (F). In case (D) the phase randomization from shot to shot was obtained by changing the path of one of the two fields at a frequency of $\sim$100 Hz with a piezoelectric device, Pz, covering a travel length of 1.28 $\mu$m. Pseudo-thermal light was obtained by means of a diffuser, D. Less than one speckle matched, in cases (B) and (E), the head of the fiber delivering the light to the detector. The insertion of lens L together with the pin-hole PH provided the coupling to the fiber of multiple speckles in cases (C) and (F).
\par\noindent
To obtain the $F_v$-value as a function of $\bar{v}$, we inserted a polarizer, P, as shown in Fig.~\ref{f:scheme}. This amounts to changing the overall detection efficiency $\eta$.
In each of the cases (A) to (F), we recorded at least 30000 $v$-values for each angle of the polarizer and repeated the measurements upon blanking either of the BS inputs.
We also measured one set of data in the absence of light, whose mean value was adopted to set the origin of the $v$-values (i.e. $v=0$) of the data collected for each light field.
Note that, using a single detector and a single laser and not moving any optical component throughout all the experiments make $\eta$ virtually constant if not for the changes in the polarizer rotation angle.
\par\noindent
For each set of data we calculated mean value, $\bar{v}$, and variance, $\sigma^{(2)}_{v}$, thus obtaining the $F_v$-value corresponding to each $P_v$ distribution, upon correcting $\sigma^{(2)}_{v}$ for the variance of the data set measured in the dark.

\begin{figure}[h]
\includegraphics[width=160mm]{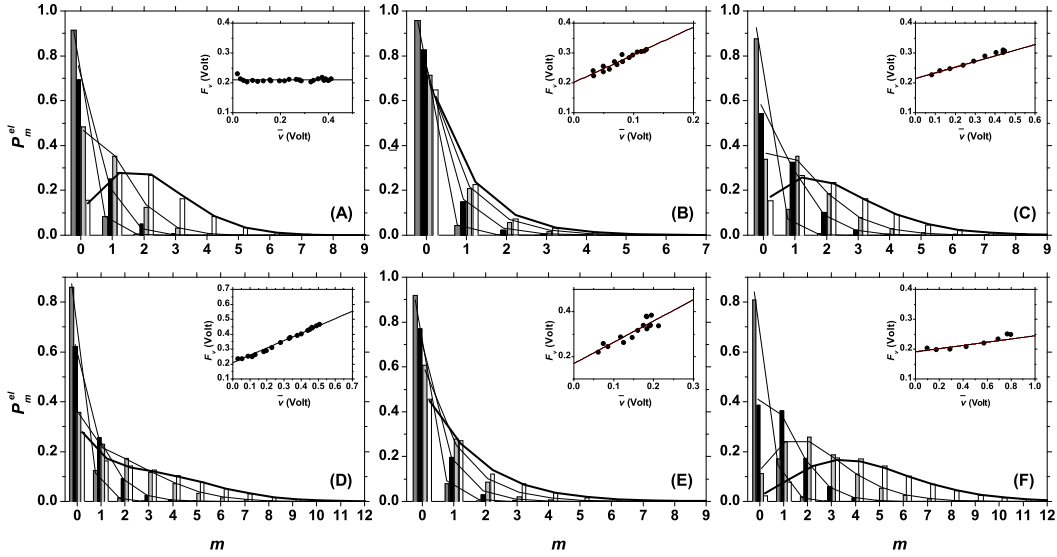}
\caption{Insets: Fano factor, $F_v$, as a function of $\bar{v}$, for the different light states. Bars: reconstructed photoelectron distributions, $P_{m}^{\ el}$, for some of the data sets used to calculate the Fano factor. Lines: theoretical curves.} \label{f:stats}
\end{figure}

The insets in Fig.~\ref{f:stats} display the plots of $F_v$ as a function of $\bar{v}$ we obtained in cases (A) to (F). In agreement with Eq.~(\ref{eq:fanoQ}) the experimental data are well fitted by straight lines, whose intercepts at $\bar{v}=0$ give $\gamma$. We find the following values of $\gamma$, in Volt: 0.21 (A), 0.20 (B), 0.21 (C), 0.21 (D), 0.18 (E) and 0.19 (F). The values of the slopes of the best fitting straight lines, which are 0 in (A), 0.980 in (B), 0.189 in (C), 0.491 in (D), 0.863 in (E) and 0.054 in (F), can be interpreted by using the results in the Section~\ref{sec:teo}. Note that in cases (A) and (B), which correspond to coherent and single-mode thermal lights, respectively, we virtually find the slopes expected, that is zero and one (see Eq.~(\ref{eq:fanoCOH}) and Eq.~(\ref{eq:fanoTHER})).

Once $\gamma$ is determined, we can reconstruct the photoelectron statistics, $P_m^{\ el}$. We convert voltages $v$ into number of photoelectrons $m$ by dividing the $v$-values by the value of $\gamma$ according to Eq.~(\ref{eq:outs}). We then rebin the obtained distributions in unitary bins to find $P_{m}^{\ el}$. In the main panels of Fig.~\ref{f:stats} the bar plots are the reconstructed photoelectron distributions for some of the $v$-data sets used to obtain the calibrations in the insets and the white-bar plots correspond to the cases of the maximum $\bar{v}$-values. For the last ones we have the following mean values: $\bar{m}\cong 1.95$ in (A), 0.60 in (B), 2.07 in (C), 2.41 in (D), 1.15 in (E) and 4.25 in (F).
The bar plots in grey/black correspond to different (lower) $\bar{m}$-values (data not shown) and equal ratios between the mixed fields.

In the case of the coherent field (A), knowing $\bar{m}$ is enough to calculate the theoretical distributions of photoelectrons by using Eq.~(\ref{eq:poiss}), which are displayed as full lines in Fig.~\ref{f:stats}~(A).
To assess the quality of the reconstruction we calculate the value of the fidelity:
\begin{equation}
 f  = \sum_{j=0}^{\infty}\sqrt{P_j^{\ el}P_j} \ .\label{eq:fidelity}
\end{equation}
For all the reconstructions in the figure we find $f=0.9996\pm 0.0004$. In particular the reconstruction of the white-bar plot having the above-mentioned $\bar m$ yields $f=0.9996$.

The case of single-mode thermal light (B), is similar. By using the experimental $\bar m$-values into Eq.~(\ref{eq:term}) we obtain the curves displayed in Fig.~\ref{f:stats}~(B) and $f=0.9993\pm 0.0004$. In particular $f=0.9992$ in the case of maximum intensity.

Cases (A) and (B) are the only ones for which the slope, $Q/\bar n$, of $F_v$ as a function of $\bar v$ identifies the statistics of the light. The proximity of the slopes of the fitting straight lines to 0 and 1, respectively, makes the high values of $f$ a result to be expected.

In case (C), in which the field is the superimposition of $\mu$ equally populated thermal modes, in order to calculate the theoretical distributions by means of Eq.~(\ref{eq:multit}) we need the additional information on the number of modes. However, this information is provided by the slope of $F_v$. For the data displayed in Fig.~\ref{f:stats}~(C) we have $\mu=5.3$. The values of the fidelity are in the range $f=0.999 \pm 0.001$ and, in particular, $f= 0.9993$ for the most intense field.

Calculating the theoretical distributions for the mixed field states in cases (D), (E) and (F), requires the further knowledge of the relative populations of the two fields. We get this information from the measurements that we repeated upon blanking either of the BS inputs.

In case (D) we determined the ratio $R = \bar{v_1}/\bar{v_2}\sim 0.927$, which remains the same for all the data in Fig.~\ref{f:stats}~(D), from the independent measurements of each of the coherent beams. By using this ratio and the experimental mean values in Eq.~(\ref{eq:phaseaver}) we calculated the theoretical curves displayed as full lines in Fig.~\ref{f:stats}~(D) and found fidelity values in the range $f=0.9995 \pm 0.0003$ and $f= 0.9992$ for the most intense field. By writing the slope $K$ in Eq.~(\ref{eq:fanoAVER}) as $K= 2R/(R+1)^2$ and using the above value of $R$, we find $K=0.494$, to be compared with the slope of the fitting straight line in the inset ($K = 0.491$).

In case (E) we measured a ratio $R = \bar{v}_{th}/\bar{v}_{\alpha}\cong 1.535$ between the thermal field and the coherent displacement field. The theoretical curves in Fig.~\ref{f:stats}~(E) were calculated by inserting this ratio and the experimental mean values in Eq.~(\ref{eq:DISPLterm}). The corresponding fidelity values are in the range $f=0.998 \pm	0.002$, being $f=0.9992$ for the most intense field. The slope $K_1$ in Eq.~(\ref{eq:fanoDISPLter}), which turns out to be $K_1= R(R+2)/(R+1)^2$, becomes $K_1=0.844$  by using the above value of $R$. Note that the slope of the fitting straight line is $K_1 = 0.863$. These results are satisfactory even if the experimental data points representing $F_v$ are rather noisy (see inset in Fig.~\ref{f:stats}~(E)). Note that also the corresponding data in the inset of Fig.~\ref{f:stats}~(B) are rather noisy, a fact that testifies the difficulty in producing a reliable single-mode thermal field.

Finally, in case (F), to calculate the theoretical distributions we need to know both the ratio $R = \bar{V}_{th}/\bar{v}_{\alpha}$ between the multi-mode thermal field and the coherent displacement field and the number of modes $\mu$. The theoretical curves in Fig.~\ref{f:stats}~(F) were calculated by using, in Eq.~(\ref{eq:DISPLmterm}), the experimental value $R= 2.446$ and setting $\mu = 8$. This value of $\mu$ optimizes the fidelity in the case of the most intense field. As a check of the choice of the parameter $\mu$, we compare the value of the slope $K_2= R/\mu(R+2\mu)/(R+\mu)^2$ in Eq.~(\ref{eq:fanoDISPLmter}) as obtained by using $R$ and $\mu$, $K_2 = 0.052$, with that evaluated from the fit of the Fano factor, $K_2 = 0.054$: the result is satisfactory, as confirmed by the fidelity values, which are in the range $f=0.9992 \pm 0.0003$, being $f=0.9990$ that of the most intense field.

\section{RECONSTRUCTION OF THE PHOTON STATISTICS}\label{sec:onoff}

It has been shown that suitably elaborating the probability of zero detected photons, $P_{0}^{\ el}$, at varying the overall detection efficiency, $\eta$, can lead to the reconstruction of the photon-number distribution $P_{n}^{\ ph}$ \cite{zambra2005}. The data needed to pursuit this approach can be collected by using a detector whose output simply allows discriminating the case in which no photon is detected ($m=0$) from any case in which $m\geq 1$. Obviously our HPD has such a feature.

\begin{figure}[h]
\includegraphics[width=160mm]{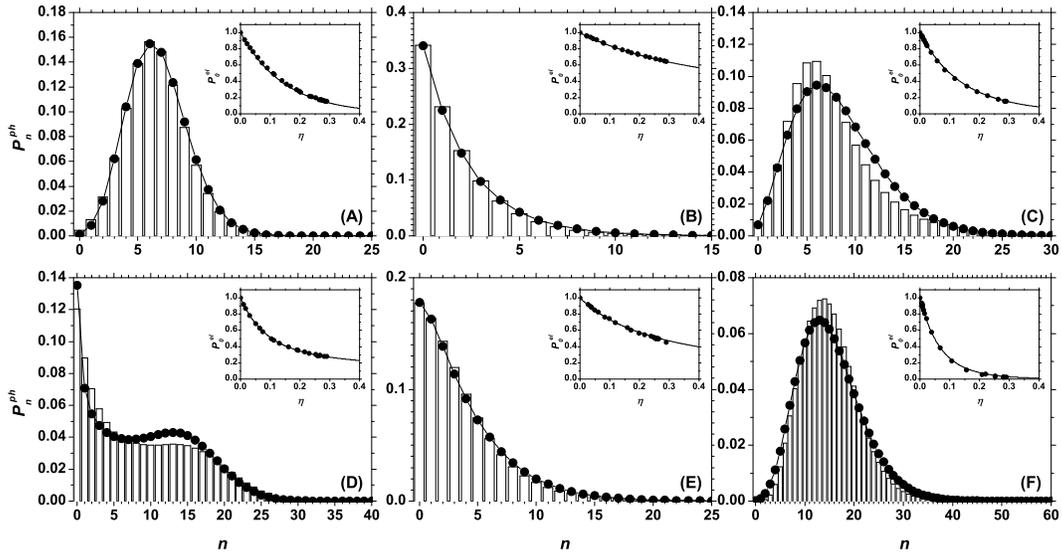}
\caption{Bars: reconstructed photon-number distributions, $P_{n}^{\ ph}$, for the different light states. The displayed reconstructions correspond to the measurements performed with $\eta=\eta_{max}$, whose $P_{m}^{\ el}$ are plotted as white bars in Fig.~\ref{f:stats}. Dots: theoretical curves, $P_n$ (the connecting lines are a guide for the eye). Insets: values of $P_{0}^{\ el}$ as a function of $\eta$ (dots) and theoretical behavior expected for each field (lines).} \label{f:phstats}
\end{figure}

We note that the data points in each of the insets in Fig.~\ref{f:stats} were obtained from measurements performed at varying $\eta$ up to $\eta_{max}$. We thus revisited our experimental data to the maximum likelihood method to recover the photon-number statistics as described in \cite{zambra2005}. To this aim, we took the values of $P_{0}^{\ el}$ from all the experimental $P_{m}^{\ el}$ distributions and assigned the corresponding values of $\eta$ as $\eta = \eta_{max} {\bar v}/{\bar v}_{max}$. The data were then processed by the maximum likelihood algorithm giving an iterative solution for $P_{n}^{\ ph}$ corresponding to $\eta_{max}$ \cite{zambra2005}. In the insets of Fig.~\ref{f:phstats} we plot $P_{0}^{\ el}$ as a function of $\eta$ and, as full line, the theoretical behavior expected for the field considered. We point out that the almost ideal dependence of our $P_{0}^{\ el}$-values on $\eta$, which can be observed in the insets, is not enough to guarantee high fidelity reconstruction \cite{zambra2005}.

In the main panels of Fig.~\ref{f:phstats} we show, as bars, the reconstructed $P_{n}^{\ ph}$ distributions that correspond to the photoelectron distributions displayed as white bars in Fig.~\ref{f:stats}. The reconstruction algorithm was stopped after a number of iterations, which was different from state to state, giving the reconstructed $P_{n}^{\ ph}$ with $\bar{n}$-value best matching the expected one, $\bar{n}_{max} = \bar{v}_{max}/(\gamma\eta_{max})$. The iterations were: 1000 in (A), 1500 in (B), 1000 in (C), 48000 (D), 1500 in (E) and 10000 in (F).

In each panel in Fig.~\ref{f:phstats} the dots show the plot of the corresponding theoretical $P_{n}$ distribution (see Section~\ref{sec:teo}) as calculated by using the experimental values of $\bar{n}_{max}$, $R$ and $\mu$ (in cases (C) and (F)).
The $f$-values evaluated in analogy with Eq.~(\ref{eq:fidelity}) for the reconstructions in Fig.~\ref{f:phstats} are: 0.9988 (A), 0.9993 (B), 0.9960 (C), 0.9975 (D), 0.9973 (E) and 0.9950 (F). It may be noted that these values are still rather close to one, but lower than those for the corresponding reconstructed $P_{m}^{\ el}$ distributions (see white-bar plots in Fig.~\ref{f:stats}).

In particular, we note that the reconstruction in cases (C), (D) and (F) is not as good as in the others. This is mainly due to the fact that cases (C), (D) and (F) represent the cases with the highest mean number of photons for which we can have fewer significant ({\it i.e.} non-zero) values of $P_{0}^{\ el}$ to be used in the reconstruction algorithm.
In addition, in case (D) the low quality of the reconstruction confirms the recognized difficulty of the method in recovering multi-peaked distributions \cite{zambraSOLO}. An improvement in this direction could be given by an increase in the overall quantum efficiency.
Moreover, in cases (C) and (F), additional problems arise from the preparation of the multi-mode thermal state to be measured, which, as described above, was generated by means of a rotating ground glass plate, selected by an aperture and focused by a lens. Any instability affecting the collection of the light by the fiber is more detrimental in this case. On the contrary, the multi-mode thermal state produced by a spontaneous down-conversion process we presented in \cite{zambra2005}, which was more robust, could be better reconstructed.
We finally note that the self-consistent method for reconstructing the photoelectron distributions $P_m^{el}$ we presented in this paper is not as much sensitive to noise and works very well also in cases (C) and (F).

%

\section*{CONCLUSIONS}

We have demonstrated that it is possible to implement a procedure to recover the distribution of detected photons that avoids calibration both of the photo-detector and of the electronics processing its output. The data to be treated are the voltages, into which the amplified charge of the photoelectrons are converted, which are the most straightforward results of this kind of measurements. The procedure employs the evaluation of the Fano factor of the voltages measured at different values of the overall detection efficiency. It has been satisfactorily tested on classical states endowed with non-trivial statistics. In all the cases we achieved high-fidelity reconstructions of the photoelectron statistics and found results in good agreement with the photon-number statistics, independently reconstructed. Works are in progress to apply the method presented here to nonclassical fields, such as squeezed states and conditionally prepared states. We expect to be able to characterize the nonclassicality of the state by intensity measurements.
%

%

\end{document}